\documentstyle[12pt]{article}
\title{
\vspace{-8mm}
\rightline{\normalsize HUB--EP--98/15}
\vspace{3mm} 
\bf On the Topological Term\\
in the String Representation of the\\
Wilson Loop in the Dilute Instanton Gas}
\author{D.V. ANTONOV \thanks{E-mail addresses: 
antonov@pha2.physik.hu-berlin.de and antonov@vxitep.itep.ru}{\,}
\thanks{On leave of absence from the Institute of Theoretical and 
Experimental Physics, B. Cheremushkinskaya 25, 117 218, Moscow, Russia.}{\,}
\thanks{
Supported by Graduiertenkolleg {\it Elementarteilchenphysik} and DFG-RFFI, 
grant 436 RUS 113/309/0.}
\\
and\\
D. EBERT \thanks{E-mail address: debert@qft2.physik.hu-berlin.de}{\,} 
\thanks{Supported by DFG-RFFI, grant 436 RUS 113/309/0.} 
\\
{\it Institut f\"ur Physik, Humboldt-Universit\"at zu Berlin,}\\
{\it Invalidenstrasse 110, D-10115, Berlin, Germany}}
\date{}
\begin{document}
\maketitle
\vspace{1cm}
\centerline{\bf {Abstract}}

\vspace{3mm}
A topological term related to the number of self-intersections of the 
string world-sheet 
is shown to emerge in the string representation of the 
Wilson loop in the dilute 
instanton gas. The coupling constant of this term occurs to be proportional 
to the topological charge of the instanton gas under consideration.

\vspace{6mm}

Recently, a string representation of the Wilson loop in the framework of 
the Method of Vacuum Correlators [1-3] has been proposed [4,5]. Within 
this approach, the expansion of the Wilson loop in powers of the derivatives 
with respect to the string world-sheet coordinates has been performed.
In the lowest orders, this yielded the Nambu-Goto and the rigidity terms 
in the effective action, whose coupling constants were expressed in 
terms of the bilocal correlator of the gauge field strength tensors. In this 
way, the bare coupling constant of the rigidity term has been obtained to be 
negative, which is important for the stability of the string [6].
However, as it has been explained in Ref. [7], if one considers this 
coupling constant as a running one, this could 
lead to the problem of crumpling of the string world-sheet in the 
infrared region. A possible way of the solution of this problem,  
proposed by the author of Ref. [7], is based on the introduction of the 
so-called topological term, which is 
equal to the algebraic number of self-intersections of the string 
world-sheet, into the string effective action. 
Then by adjusting the coupling constant of this term, one 
can eventually 
arrange the cancellation of contributions into the string partition function 
coming from highly crumpled surfaces, whose intersection numbers differ  
by one from each other. 

Thus it looks natural to address the problem 
of derivation of the topological term 
from the Lagrangians of various gauge theories, 
which possess confining phases and therefore admit string representations. 
A progress in this direction has been recently achieved in Ref. [8] 
for the case of 4D compact QED with a $\theta$-term. In the dual 
formulation of the Wilson loop in this theory, the latter one occured to 
be crucial for the formation of the topological string term. 
However, such 
a mechanism of generation of a topological term is difficult to 
work out in the non-Abelian case, e.g. in gluodynamics, due to 
our inability to construct the exact dual formulation of the Wilson loop in 
this theory. Therefore, it looks suggestive to search for some 
model of the gluodynamics vacuum, which might lead to the 
appearance of the topological term in the string representation of the 
Wilson loop in this theory. In the present paper, we shall elaborate out    
one of such possibilities. To this end, we shall make use of  
recent results concerning the calculation of the         
field strength correlators in the dilute 
instanton gas model [9]. There it 
has been demonstrated that for the case of an instanton gas with broken 
$CP$ invariance, the bilocal field strength correlator contains 
a term proportional to the tensor $\varepsilon_{\mu\nu\lambda
\sigma}$. This term is absent in the case of a $CP$-symmetric vacuum, 
since it is proportional to the topological charge of the system, 
$V\left(n_4-\bar n_4\right)$, 
where $V$ is the four-volume of observation, and within the notations of 
Ref. [9],  
$n_4$ and $\bar n_4$ stand for the densities 
of instantons and antiinstantons ($I$'s and $\bar I$'s for short),  
respectively. 
Similarly, we shall also work in the approximation of a 
dilute $I-\bar I$ gas with  
fixed equal sizes $\rho$ of $I$'s and $\bar I$'s.  

Let us now briefly remind the main idea of Ref. [4]. According to this 
paper, the nonlocal 
string effective action associated with the surface of a minimal area 
$S_{\rm min.}$, 
bounded by the contour of the Wilson loop, is defined as 
${\cal A}_{\rm eff.}=-\ln \left<W(S_{\rm min.})\right>$. Here the Wilson loop 
itself 
could be written within the Method of Vacuum Correlators [1-3] as follows 

$$\left<W(S_{\rm min.}
)\right>=\frac{1}{N_c^2-1}{\rm tr}\exp\left(-\int\limits_{S_{\rm 
min.}}^{}d\sigma_{\mu\nu}(x)\int\limits_{S_{\rm min.}}^{}d\sigma_{\lambda
\sigma}(x')\left<F_{\mu\nu}(x,x_0)F_{\lambda\sigma}(x',x_0)\right>\right), 
\eqno (1)$$
where $F_{\mu\nu}(x,x_0)$ is the so-called shifted strength 
tensor of the gauge field, which is related to the usual tensor 
$F_{\mu\nu}(x)\equiv F_{\mu\nu}^a(x)T_{\rm adj.}^a$ as follows  

$$F_{\mu\nu}(x,x_0)\equiv\Phi(x_0,x)F_{\mu\nu}(x)\Phi(x,x_0).$$
In this equation, 
$\Phi(x,x_0)$ stands for the parallel transporter factor of the 
gauge field defined on the straight-line contour, which goes from  
the fixed point $x_0$ to the point $x$. The bilocal 
correlator of the field strength tensors standing on the R.H.S. of Eq. (1)
can be further parametrized by the two independent Lorentz structures, 
whose coefficient functions denoted in Refs. [1-3] as $D$ and $D_1$ 
contain the main information about 
both perturbative and nonperturbative properties of the gluodynamics 
vacuum. These two functions decrease 
fastly at some distance,  
which in Refs. [1-3] has been called the 
correlation length of the vacuum and according to the lattice data [10] 
is equal for the $SU(3)$-case to 
$0.2{\,} {\rm fm}$. Then, since the typical size $r$ of the 
contour of the Wilson loop is of the order of $1.0{\,} {\rm fm}$ [11] in the 
confining regime under study, 
one can perform 
the expansion of the nonlocal action ${\cal A}_{\rm eff.}$ in powers 
of $\left(T_g/r\right)^2$. 
This expansion is in fact 
nothing else 
but the expansion in powers of the derivatives with respect to the string 
world-sheet coordinates, mentioned in the beginning of the present paper.

The new structure arising in the bilocal correlator in the 
$I-\bar I$ gas reads [9]

$$\Delta{\,}{\rm tr}\left<F_{\mu\nu}(x, x_0)F_{\lambda\sigma}
(x', x_0)\right>=
8\left(n_4-\bar n_4\right)I_r\left(\frac{(x-x')^2}{\rho^2}
\right)\varepsilon_{\mu\nu\lambda\sigma}. \eqno (2)$$
In Eq. (2), the asymptotic behaviour of the function $I_r\left(z^2\right)$ 
at $z\ll 1$ and $z\gg 1$ has the following form  

$$I_r\left(z^2\right)\longrightarrow \frac{\pi^2}{6}, \eqno (3)$$
and 

$$I_r\left(z^2\right)\longrightarrow\frac{2\pi^2}{\left(z^2\right)^2}
\ln z^2, \eqno (4)$$
respectively. 

In what follows, we are going to present the leading term in the 
derivative expansion of the correction 
to the nonlocal string effective action
 
$$\Delta{\cal A}_{\rm eff.}=-\ln \Delta\left<W(S_{\rm min.})\right>, 
\eqno (5)$$ 
where 
$\Delta\left<W(S_{\rm min.})\right>$ is a correction to the 
expression (1) for the Wilson loop, following from Eq. (2) in the 
$CP$-broken vacuum. 
Here we shall not 
be interested in the corrections to the Nambu-Goto and rigidity terms 
arising due to 
additional contributions from the $I-\bar I$ gas 
to the functions $D$ and $D_1$, which stand at Lorentz structures without 
the tensor $\varepsilon_{\mu\nu\lambda\sigma}$. 
According to Ref. [4], this could be easily done by carrying out the 
corresponding integrals of the functions $D$ and $D_1$ in this gas. 
Notice only that, as it has already been mentioned in Ref. [9], 
due to the reasons discussed in details in Refs. [2] and [3], 
a correction to the string tension of the Nambu-Goto term obtained 
in such a way from the $I-\bar I$ gas contribution to the function 
$D$ should be cancelled by the contributions coming 
from the higher cumulants in this gas.

Let us now turn ourselves to the expansion of the 
correction (5), emerging from the term (2), in powers of 
the derivatives with respect 
to the string world-sheet coordinates. Firstly, one can see that 
since $\tilde t_{\mu\nu}t_{\mu\nu}=0$, where $t_{\mu\nu}$ is the 
extrinsic curvature tensor of the string world-sheet, an analog of the 
Nambu-Goto term for this correction vanishes. Then, similarly to Ref. [4],  
we get  

$$\Delta{\cal A}_{\rm eff.}=
\alpha\nu+{\cal O}\left(\frac{\rho^6\left(n_4-\bar n_4\right)}
{r^2}\right), $$
where 

$$\nu\equiv\frac{1}{4\pi}\varepsilon_{\mu\nu\lambda\sigma}
\int d^2\xi\sqrt{g}g^{ab}(\partial_a t_{\mu\nu})(\partial_b 
t_{\lambda\sigma}) $$ 
is the algebraic number of self-intersections of the string world-sheet, and

$$\alpha=16\pi\rho^4\left(n_4-\bar n_4\right)\int d^2zz^2 I_r\left(
z^2\right) \eqno (6)$$
is the corresponding coupling constant. Here $g^{ab}$ stands for the induced 
metric tensor of the string world-sheet, whose determinant is denoted by $g$. 

Note that 
the averaged separation 
between the nearest neighbors in the $I-\bar I$ gas  
is given by $R=\left(n_4+\bar n_4\right)^{-\frac14}$. 
According to phenomenological considerations one obtains for the 
$SU(3)$-case, $\rho/R\simeq 1/3$ [12]; 
see also Ref. [13], where the ratio $\rho/R$ 
has been obtained from direct 
lattice measurements to be $0.37-0.40$. 
$R$ should then serve as a distance cutoff in the 
integral standing on the R.H.S. of Eq. (6). 
Taking this into account, one gets from Eqs. (3), (4), and (6)   
the following approximate value of $\alpha$ 

$$\alpha\simeq(2\pi\rho)^4\left(n_4-\bar n_4\right)\left[\frac{1}{12}+
\left(\ln\frac{R^2}{\rho^2}\right)^2\right], \eqno (7)$$
where the second term in the square brackets on the R.H.S. of Eq. (7), 
emerging due to Eq. (4),  
is much larger than the first one, emerging due to Eq. (3).

In conclusion, we have found that in the $I-\bar I$ gas with a 
nonzero topological charge, there appears a topological term in the 
string representation of the Wilson loop. The coupling 
constant of this term is given by Eq. (7). Together with the Nambu-Goto and 
rigidity terms obtained in Ref. [4], this term forms the effective 
Lagrangian of the gluodynamics string following from the Method of 
Vacuum Correlators. This string is stable and eventually free of 
crumpling.

\vspace{6mm}
{\large\bf Acknowledgments}

\vspace{3mm}
D.A. would like to thank the theory group 
of the Quantum Field Theory Department of the Institut f\"ur 
Physik of the Humboldt-Universit\"at of Berlin for kind 
hospitality.

\vspace{6mm}
{\large\bf References}

\vspace{3mm}
\noindent
$[1]$~H.G. Dosch, Phys. Lett. B {\bf 190}, 177 (1987); Yu.A. Simonov, 
Nucl. Phys. B {\bf 307}, 512 (1988); H.G. Dosch and Yu.A. Simonov,
Phys. Lett. B {\bf 205}, 339 (1988), Z. Phys. C {\bf 45}, 147 (1989); 
Yu.A. Simonov, Nucl. Phys. B {\bf 324}, 67 (1989), 
Phys. Lett. B {\bf 226}, 151 (1989), Phys. Lett. B {\bf 228}, 413 (1989), 
Sov. J. Nucl. Phys. {\bf 54}, 115 (1991).\\
$[2]$~Yu.A. Simonov, Sov. J. Nucl. Phys. {\bf 50}, 310 (1989).\\
$[3]$~Yu.A. Simonov, Phys. Usp. {\bf 39}, 313 (1996).\\
$[4]$~D.V. Antonov, D. Ebert, and Yu.A. Simonov, Mod. Phys. Lett. A
{\bf 11}, 1905 (1996).\\ 
$[5]$~D.V. Antonov and D. Ebert, Mod. Phys. Lett. A {\bf 12}, 2047 
(1997).\\
$[6]$~H. Kleinert, Phys. Lett. B {\bf 211}, 151 (1988); K.I. Maeda and 
N. Turok, Phys. Lett. B {\bf 202}, 376 (1988); S.M. Barr and D. Hochberg, 
Phys. Rev. D {\bf 39}, 2308 (1989); H. Kleinert, Phys. Lett. 
B {\bf 246}, 127 (1990); P. Orland, Nucl. Phys. B {\bf 428}, 221 (1994); 
H. Kleinert and 
A.M. Chervyakov, preprint hep-th/9601030 (1996) 
(in press in Phys. Lett. B).\\ 
$[7]$~A.M. Polyakov, Nucl. Phys. B {\bf 268}, 406 (1986).\\
$[8]$~M.C. Diamantini, F. Quevedo, and C.A. Trugenberger, Phys. Lett. 
B {\bf 396}, 115 (1997).\\
$[9]$~E.-M. Ilgenfritz, B.V. Martemyanov, S.V. Molodtsov, 
M. M\"uller-Preu{\ss}ker, and Yu.A. Simonov, preprints 
HUB-EP 97/97, KANAZAWA 97-22, ITEP-PH-15-97, and hep-ph/9712523 
(1997).\\
$[10]$A. Di Giacomo and H. Panagopoulos, Phys. Lett. B {\bf 285}, 133 
(1992).\\ 
$[11]$I.-J. Ford et al., Phys. Lett. B {\bf 208}, 286 (1988); 
E. Laermann et al., Nucl. Phys. B (Proc. Suppl.) {\bf 26}, 268 (1992).\\ 
$[12]$E. Shuryak, Nucl. Phys. B {\bf 203}, 93, 116, 140 (1982).\\
$[13]$M.-C. Chu, J. Grandy, S. Huang, and J. Negele, Phys. Rev. Lett. 
{\bf 70}, 225 (1993); Phys. Rev. D {\bf 49}, 6039 (1994).

\end{document}